\documentclass[12pt,a4paper]{article}
\usepackage{latexsym}
\usepackage{amssymb}
\usepackage{amsthm}
\usepackage{amsmath}
\usepackage[cp1251]{inputenc}

\newcommand{\ignore}[1]{}

\newtheorem*{theorem*}{Theorem}

\newtheorem*{claim*}{Claim}

\renewcommand{\SS}{\mathbb S}
\newcommand{\e}{\circ}

\renewcommand{\proof}{\noindent $\rhd$\;}

\begin{document}
\title{A note on the depth of optimal fanout-bounded prefix circuits}
\date{}
\author{I. S. Sergeev\footnote{e-mail: isserg@gmail.com}}
\maketitle

\begin{abstract}
It is shown that the minimal depth of an optimal prefix circuit
(i.e., a zero-deficiency circuit) on $N$ inputs with fanout
bounded by $k$ is ${\log_{\alpha_k} N \pm O(1)}$, where $\alpha_k$
is the unique positive root of the polynomial ${2+x+ x^2+\ldots +
x^{k-2}-x^k}$. This bound was previously known in the cases $k=2$
and $k=\infty$.
\end{abstract}

{\bf Introduction}

\medskip

Let $(\SS,\,\e)$ be a semigroup. The set of functions
\begin{equation}\label{pref}
s_i = x_1 \e x_2 \e \ldots\e x_{i}, \qquad 1\le i \le N,
\end{equation}
is called the system of {\it prefix sums} of variables
$x_1,\ldots,x_N$ taking values in~$\SS$. Circuits of functional
elements over the basis $\{x\e y,\, x\}$ that implement the
system~(\ref{pref}) are called {\it prefix circuits}. The
number~$N$ (of circuit inputs) is called the {\it width} of a
circuit. By the {\it complexity} of a circuit we will (as usual)
mean the total number of binary elements ``$\e$'' in it. The need
for identity elements appears only when the circuit fanout is
bounded. The {\it depth} of a circuit is the maximum number of
elements (of both types) in an input-output path.

We consider {\it universal} prefix circuits that correctly compute
sums regardless of the choice of a semigroup $\SS$. It is easy
to verify that in a minimal (i.e., not containing elements
unconnected to outputs) universal circuit, only interval sums are
computed via operations of the form ${p_1 \e p_2}$, where $p_1 =
{x_i \e x_{i+1} \e \ldots \e x_j}$ and $p_2 = {x_{j+1} \e x_{j+2}
\e \ldots \e x_l}$. If a node in the circuit computes the sum $x_i
\e \ldots \e x_j$, then $j$ is called the {\it index} of this
node.

Obviously, all sums $s_i$ can be computed sequentially, with a
minimum number of $N-1$ operations~``$\e$''. The complexity~$C$ and
the depth~$D$ of a prefix circuit of width $N$ are related as $C+D
\ge 2N-2$~\cite{fic83e,sni86e}, so the complexity of parallel prefix
circuits cannot be significantly less than $2N$. Prefix circuits
for which the equality $C+D = 2N-2$ holds are called {\it optimal}
or circuits with zero deficiency.

As is known, an optimal prefix circuit of depth $D$ on $N$ inputs
(when it exists) has the following structure. Its elements either
belong to the {\it framework tree} of depth $D$, which is the
subcircuit computing the sum of all inputs, or are outputs of the
circuit. Each of these two sets of elements has cardinality $N-1$,
but $D$~elements of the {\it principal chain}, i.e., the chain
connecting the first input with the last output of the circuit,
belong to both sets\footnote{Moreover, upon transposition, i.e.,
reversing the direction of the circuit, both sets are mapped
into each other.}. Therefore, a circuit has complexity $2N-D-2$.
For more details, see, e.g.,~\cite{zcg06e}.

Let $D(N,k)$ denote the minimum possible depth of an optimal
circuit on $N$~inputs with fanout bounded by~$k$.

It is shown in~\cite{zcg06e} that
\begin{equation}\label{zcg}
 D(N,\infty) = d = \log_{\varphi} - O(1) \approx 1.44\log_2 N - O(1),
\end{equation}
where $\Phi_{d+3}$ is the nearest number from the Fibonacci
sequence $\{\Phi_m\}$ to $N+1$ from above, and $\varphi =
\frac{1+\sqrt5}2$. From~\cite{lh09e,sp06e} it follows that $D(N,2) =
\lfloor \log_2 N \rfloor + \lfloor \log_2(2N/3)\rfloor$. Exact or
at least asymptotic closed-form estimates for $D(N,k)$, where $2 <
k < \infty$, have apparently not yet been obtained, despite the
fact that, for example, in~\cite{sp06e} optimal fanout-bounded
circuits of extreme width were constructed.

Let $\alpha_k$ denote the unique positive root of the polynomial
$P_k(x) = {2+x+ x^2+\ldots + x^{k-2}-x^k}$. Further, we will prove
\begin{theorem*}
$D(N,k) = \log_{\alpha_k} N \pm O(1)$.
\end{theorem*}

It is easy to verify that $\alpha_k \to \frac{1+\sqrt5}2$ as $k
\to \infty$, which is consistent with~(\ref{zcg}). In particular, the theorem
implies $D(N,3) \sim 1.65\ldots \cdot \log_2 N$, $D(N,4) \sim
1.54\ldots \cdot \log_2 N$ and already $D(N,9) \lesssim 1.45
\log_2 N$.

\medskip

{\bf Proof of the theorem}

\medskip

Consider an optimal prefix circuit of depth $D$ with $N$ inputs.
Let its principal chain be formed by a sequence of nodes $v_0,
v_1, \ldots, v_D$, where $v_0$ coincides with input $x_1$ and an
arbitrary node $v_d$ is located at depth $d$.

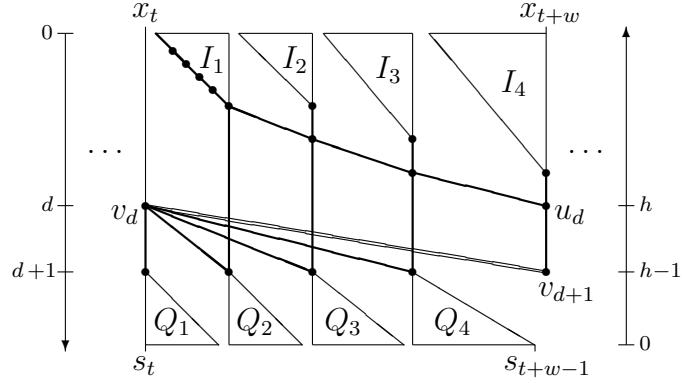
\begin{figure}[htb]
\begin{center}
\begin{picture}(250,137)(0,-7)

\thinlines


\put(20,120){\vector(0,-1){120}}
\multiput(17,30)(0,25){2}{\line(1,0){6}}
\put(17,120){\line(1,0){6}}

\put(230,2.5){\vector(0,1){120}}
\multiput(227,30)(0,25){2}{\line(1,0){6}}
\put(227,2.5){\line(1,0){6}}


\multiput(81.25,2.5)(31.25,0){2}{\line(0,1){117.5}}
 \put(150,2.5){\line(0,1){117.5}}
 \put(200,30){\line(0,1){92.5}}

 \put(50,0){\line(0,1){122.5}}


\multiput(50,30)(31.25,0){3}{\circle*{3}}
 \put(200,30){\circle*{3}}
 \put(150,30){\circle*{3}}
\multiput(50,55)(150,0){2}{\circle*{3}}
\multiput(150,67.5)(50,0){2}{\circle*{3}}
\multiput(150,80)(-37.5,0){2}{\circle*{3}}
\multiput(81.25,92.5)(31.25,0){2}{\circle*{3}}

\multiput(75.25,98.5)(-5,5){4}{\circle*{3}}

\multiput(81.25,92.5)(31.25,0){2}{\line(-1,1){27.5}}
\multiput(81.25,120)(31.25,0){2}{\line(-1,0){27.5}}
 \put(150,80){\line(-5,6){33.33}}
 \put(150,120){\line(-1,0){33.33}}
 \put(200,67.5){\line(-5,6){43.75}}
 \put(200,120){\line(-1,0){43.75}}
\multiput(50,30)(31.25,0){2}{\line(1,-1){27.5}}
\multiput(50,2.5)(31.25,0){2}{\line(1,0){27.5}}
 \put(112.5,30){\line(5,-4){34.375}}
 \put(112.5,2.5){\line(1,0){34.375}}
 \put(150,30){\line(5,-3){45.83}}
 \put(150,2.5){\line(1,0){45.83}}

 \put(195.83,2.5){\line(0,-1){2.5}}


\multiput(50,54.5)(0,1){2}{\line(6,-1){150}}

\thicklines

\put(81.25,92.5){\line(-1,1){27.3}}

 \put(50,55){\line(0,-1){25}}
 \put(50,55){\line(5,-4){31.25}}
 \put(50,55){\line(5,-2){62.5}}
 \put(50,55){\line(4,-1){100}}

 \put(200,55){\line(-4,1){50}}
 \put(150,67.5){\line(-3,1){37.5}}
 \put(112.5,80){\line(-5,2){31.25}}

\multiput(81.25,30)(31.25,0){2}{\line(0,1){62.5}}
 \put(150,30){\line(0,1){50}}
 \put(200,30){\line(0,1){37.5}}


 \put(235,0.5){$\scriptstyle 0$}
 \put(235,28){$\scriptstyle h-1$}
 \put(235,53){$\scriptstyle h$}

 \put(11,118){$\scriptstyle 0$}
 \put(11,53){$\scriptstyle d$}
 \put(0,28){$\scriptstyle d\,+1$}

 \put(37,50){$v_d$} \put(197,21){$v_{d+1}$} \put(203,50){$u_{d}$}

 \put(45,126){$x_t$}  \put(190,126){$x_{t+w}$}
 \put(45,-7){$s_t$}  \put(184,-7){$s_{t+w-1}$}

 \put(70,107){$I_1$} \put(101,107){$I_2$} \put(136,103){$I_3$}  \put(183,98){$I_4$}
 \put(53,8){$Q_1$} \put(84,8){$Q_2$} \put(117,8){$Q_3$}  \put(158,8){$Q_4$}

 \put(28,75){$\ldots$} \put(208,75){$\ldots$}

\end{picture}
\caption{Structure of a segment of an optimal prefix
circuit}\label{pic_s}
\end{center}
\end{figure}

The nodes of the principal chain naturally partition the circuit
into segments. If the sums $s_t$ and $s_{t+w}$, respectively, are
calculated at nodes $v_d$ and $v_{d+1}$, then the $d$-th segment
includes the inputs and nodes with indices in the interval $[t,\,
t+w]$. The parameter $w$ denotes the segment width. The structure
of a segment of an optimal circuit is shown in Fig.~\ref{pic_s}
(the notation is standard, see, e.g.,~\cite{sp06e,zcg06e}). There
$h=D-d$.

The segment's construction is determined by two trees-subcircuits:
a binary tree directed from the inputs $x_{t+1}, \ldots, x_{t+w}$
to the root $u_d$, and a tree consistent with it, directed from
the root $v_d$ to the outputs $s_t, \ldots, s_{t+w-1}$. The fanout
of the second tree is bounded by $k$. Tree consistency means that
the second tree employs exactly the interval sums calculated by
the first tree. In particular, all descendant neighbors of the
node $v_d$ receive second inputs strictly from nodes in the chain
connecting $x_{t+1}$ and $u_d$.

The structure of one segment is independent of the structure of
the other segments. Therefore, the maximum width of a circuit of a
given depth and fanout is the sum of the maximum possible widths
of the segments.

Let $w_k(d,h)$ denote the maximum width of a pair of consistent
trees, the first of which has depth $\le d$ and the second has
depth $\le h$ and fanout bounded by~$k$. By $w_k(D)$ we denote the
maximum width of an optimal depth-$D$ fanout-$k$ circuit. We
introduce the notation
\[ w^*_k(D) = \sum_{d=0}^D w_k(d,D-d). \]
Note that
\begin{equation}\label{wlow}
w^*_k(D-1) \le w_k(D) \le w^*_k(D).
\end{equation}
The upper bound describes the maximum width of circuits in which
fanout $k+1$ is allowed as an exception for the nodes $v_d$ of the
principal chain. The lower bound describes the width of circuits
in which the fanout of nodes $v_d$ is bounded by two. In
particular, $w_2(D) = w_2^*(D-1)$.

\begin{claim*}
Let $d,h >0$ and $l=\min\{d,k-1\}$. Then
\begin{equation}\label{wk}
 w_k(d,h) = \sum_{i=1}^{l-1} w_k(d-i,h-1) + 2w_k(d-l,h-1).
\end{equation}
\end{claim*}

\proof We continue using Fig.~\ref{pic_s} as an illustration. Let
it depict a pair of consistent trees $I$ and $Q$ with root nodes
$u_d$ and $v_d$, respectively. The immediate descendants of node
$v_d$ determine a partition of the index interval $[t,\,t+w]$ into
subintervals defined by the indices of the nodes of the chain
connecting $x_{t+1}$ and $u_d$, and also a partition of both trees
into pairs of consistent subtrees $(I_j,\,Q_j)$. Consequently,
\begin{equation}\label{wk0}
 w_k(d,h) = w_k(d_1,h-1) + \ldots + w_k(d_r,h-1),  \quad
 d > d_1 > d_2 > \ldots > d_{r-1} \ge d_r.
\end{equation}

\begin{figure}[htb]
\begin{center}
\begin{picture}(247,115)(6,-5)


\thinlines


 \put(10,10){\line(1,0){60}}
 \put(10,7.5){\line(0,1){47.5}}
 \put(70,10){\line(-4,3){60}}

\multiput(42.5,68)(32.5,0){2}{\line(0,1){30}}
\multiput(42.5,68)(32.5,0){2}{\line(-1,1){27.5}}
\multiput(42.5,95.5)(32.5,0){2}{\line(-1,0){27.5}}


 \put(10,55){\circle*{3}}
 \put(75,55){\circle*{3}}
 \put(75,68){\circle*{3}}
 \put(42.5,68){\circle*{3}}
 \put(42.5,30.63){\circle*{3}}


\thicklines

 \put(75,55){\line(0,1){13}}
 \put(75,55){\line(-5,2){32.5}}

\thinlines

\multiput(42.5,12.6)(0,3){18}{\circle*{1}}


 \put(42.5,10){\line(0,-1){2.5}}
 \put(70,10){\line(0,-1){2.5}}
 \put(15,98){\line(0,-1){2.5}}

 \put(51,74){$I$} \put(25,19){$Q$} \put(46,32){$z$} \put(34,62){$y$} \put(68,47){$u$}  \put(12,57){$v$}

 \put(11,102){$x_1$}  \put(38,102){$x_{\tau}$}  \put(70,102){$x_w$}
 \put(22,100){$\cdots$} \put(52,100){$\cdots$}

 \put(6,0){$s_0$}  \put(38,0){$s_{\tau}$} \put(65,0){$s_{w-1}$}
 \put(20,-2){$\cdots$} \put(49,-2){$\cdots$}



\multiput(203,-5)(0,8){15}{\line(0,1){4}}

\thicklines

 \put(100,55){\vector(1,0){35}}

\thinlines


 \put(160,10){\line(1,0){30}}
 \put(160,7.5){\line(0,1){47.5}}
 \put(160,55){\line(4,-3){30}}
 \put(190,7.5){\line(0,1){25}}

 \put(212.5,34.4){\line(1,0){27.5}}
 \put(212.5,31.9){\line(0,1){23.1}}
 \put(240,34.4){\line(-4,3){27.5}}

\multiput(192.5,68)(52.5,0){2}{\line(0,1){30}}
\multiput(192.5,68)(52.5,0){2}{\line(-1,1){27.5}}
\multiput(192.5,95.5)(52.5,0){2}{\line(-1,0){27.5}}


 \put(160,55){\circle*{3}}
 \put(212.5,55){\circle*{3}}
 \put(245,68){\circle*{3}}
 \put(192.5,68){\circle*{3}}


 \put(240,34.4){\line(0,-1){2.5}}
\multiput(165,95.5)(52.5,0){2}{\line(0,1){2.5}}

 \put(182,83){$I_1$} \put(234,83){$I_2$} \put(169,22){$Q_1$} \put(214,38){$Q_2$}

 \put(216,56.4){$z$} \put(184,61){$y$} \put(162,57){$v$} 

 \put(159,102){$x_1$}  \put(187,102){$x_{\tau}$}  \put(205.5,102){$x_{\tau+1}$}  \put(243,102){$x_w$}
 \put(170,100){$\cdots$} \put(227.5,100){$\cdots$}

 \put(154,0){$s_0$}  \put(181,0){$s_{\tau-1}$}  \put(208,24.4){$s_{\tau}$} \put(235,24.4){$s_{w-1}$}
 \put(165,-2){$\cdots$} \put(219,22.4){$\cdots$}

\end{picture}
\caption{Transformation of a pair of consistent
trees}\label{pic_iq}
\end{center}
\end{figure}
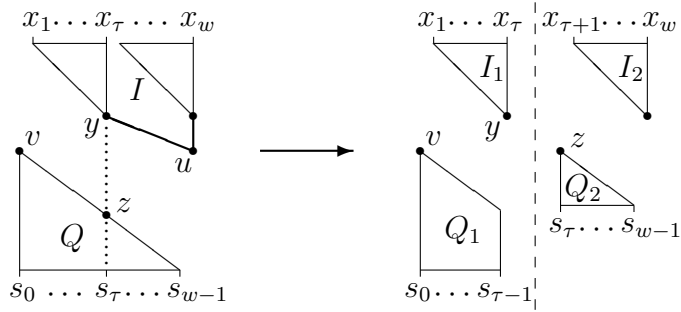

Obviously, for any $d, h$,
\begin{equation}\label{w1}
w_k(d,h) \le w_k(d+1,h). 
\end{equation}
Let us check that for $d \ge 1$ and any $h$, we also have
\begin{equation}\label{w2}
\quad w_k(d,h) \le 2w_k(d-1,h).
\end{equation}
The argument is illustrated in Fig.~\ref{pic_iq}. Consider a pair
of consistent trees of width $w = w_k(d,h)$, consisting of a
binary tree $I$ of depth $\le d$ and a $k$-ary tree $Q$ of depth
$\le h$. Let $y$ denote the closest ancestor node of root $u$ of
tree $I$ lying on the path from the first input $x_1$. Let subtree
$I_1$, rooted at $y$, have width $\tau$. Let $I_2$ denote the
subtree of tree~$I$ whose leaves are the remaining $w-\tau$
inputs. By the consistency property, the second tree $Q$ contains
a node $z$ that is an ancestor of exactly $w-\tau$ higher outputs.
Let $Q_2$ denote the subtree rooted at node $z$, and $Q_1$ denote
the tree obtained from~$Q$ by removing subtree~$Q_2$.

By construction, the pairs of trees $(I_1, Q_1)$ and $(I_2, Q_2)$ are consistent. The total width of the pairs is $w$.
Moreover, the depth of trees $I_1$, $I_2$ does not exceed $d-1$, 
and the depth of trees $Q_1$, $Q_2$ does not exceed $h$.
Hence, (\ref{w2}) is proved.

Now (\ref{wk}) immediately follows from~(\ref{wk0}) by applying rules~(\ref{w1}),~(\ref{w2}) 
and taking into account $r \le k$ and $d_r \ge 0$.
\qed

\smallskip

We proceed directly to the proof of the theorem.
Let us estimate $w_k^*(D)$. In view of (\ref{wk}), for $D \ge k$ we have
\begin{equation*}
w^*_k(D) = w_k(D,0) + \sum_{i=2}^{k-1} w^*_k(D-i) + 2w^*_k(D-k) + w_k(0,D)+\sum_{i=2}^{k-1} w_k(0,D-i).
\end{equation*}
Since $w(0,h) = w(d,0) = 1$ for any $d,h \ge 0$, we obtain the recurrence relation
\begin{equation*}
w^*_k(D) = \sum_{i=2}^{k-1} w^*_k(D-i) + 2w^*_k(D-k) + k.
\end{equation*}
This relation, given the initial values $w^*_k(0), \ldots, w^*_k(k-1)$, is resolved 
in the standard way as $w^*_k(D) \sim {c\cdot\alpha_k^D}$, where $c$ is some constant, 
since $\alpha_k$ has the largest absolute value among the roots of polynomial $P_k(x)$: 
indeed, the modulus of an arbitrary root $x$ satisfies the inequality
\[ |x|^k \le 2 + |x| + |x|^2 + \ldots + |x|^{k-2}, \]
whence $|x| \le \alpha_k$. The assertion of the theorem now follows from (\ref{wlow}).

\end{document}


\title{Замечание о глубине оптимальных префиксных схем ограниченного ветвления}
\date{}
\author{И. С. Сергеев\footnote{e-mail: isserg@gmail.com}}
\maketitle

\begin{abstract}
Показано, что минимальная глубина оптимальной префиксной схемы (т.\,е. схемы
с нулевой избыточностью) на $N$ входах с ограничением $k$ на ветвление элементов
равна ${\log_{\alpha_k} N \pm O(1)}$, где $\alpha_k$~--- единственный положительный корень
многочлена ${2+x+ x^2+\ldots + x^{k-2}-x^k}$. 
Указанная оценка ранее была известна в случаях $k=2$ и $k=\infty$.
\end{abstract}

{\bf Введение}

\medskip

Пусть $(\SS,\,\e)$~--- полугруппа. Множество функций
\begin{equation}\label{pref}
 s_i = x_1 \e x_2 \e \ldots\e x_{i}, \qquad 1\le i \le N,
\end{equation}
называется системой {\it префиксных сумм} переменных
$x_1,\ldots,x_N$, принимающих значения в~$\SS$. Схемы из
функциональных элементов над базисом $\{x\e y,\, x\}$, реализующие
систему~(\ref{pref}), называют {\it префиксными схемами}.
Число~$N$ (входов схемы) называется {\it шириной} схемы. Под {\it
сложностью} схемы будем (как обычно) понимать общее число бинарных
элементов $\circ$ в ней. Потребность в тождественных элементах
возникает только при ограничении ветвления элементов. {\it
Глубиной} схемы называется максимальное число элементов (обоих
видов) в пути, соединяющем вход и выход схемы.

Мы рассматриваем {\it универсальные} префиксные схемы, которые
правильно вычисляют суммы независимо от выбора полугруппы $\SS$.
Легко проверить, что в минимальной (т.\,е. не содержащая элементы,
не связанные с выходами) универсальной схеме вычисляются только
интервальные суммы посредством операций вида ${p_1 \e p_2}$, где
$p_1 = {x_i \e x_{i+1} \e \ldots \e x_j}$ и $p_2 = {x_{j+1} \e
x_{j+2} \e \ldots \e x_l}$. Если в вершине схемы вычисляется сумма
$x_i \e \ldots \e x_j$, то $j$ назовем {\it индексом} этой
вершины.

Очевидно, все суммы $s_i$ можно вычислить последовательно, с минимальным числом $N-1$ операций~$\e$.
Сложность~$C$ и~глубина~$D$ префиксной схемы ширины $N$ связаны соотношением
$C+D \ge 2N-2$~\cite{fic83,sni86}, поэтому сложность параллельных
префиксных схем не может быть существенно меньше~$2N$.
Префиксные схемы, для которых выполняется равенство $C+D = 2N-2$
называются {\it оптимальными} или схемами с нулевой избыточностью.

Как известно, оптимальная префиксная схема глубины $D$ на $N$
входах (когда она существует) имеет следующее строение. Ее
элементы либо принадлежат {\it каркасному дереву} глубины $D$~---
подсхеме, вычисляющей сумму всех входов,~--- либо являются
выходами схемы. Множество элементов каждого из этих двух типов
имеет мощность $N-1$, но $D$ элементов {\it главной цепочки}~---
цепи, соединяющей младший вход со старшим выходом схемы,~---
принадлежат обоим множествам\footnote{Причем при транспонировании,
т.\,е. изменении ориентации схемы, оба множества переходят друг в
друга.}. Таким образом, схема имеет сложность $2N-D-2$. Подробнее
см., например, в~\cite{zcg06}.

Пусть $D(N,k)$ означает минимально возможную глубину оптимальной схемы на $N$ входах
с ограничением $k$ на ветвление элементов.

В работе~\cite{zcg06} показано, что
\begin{equation}\label{zcg}
 D(N,\infty) = d = \log_{\varphi} - O(1) \approx 1.44\log_2 N - O(1),
\end{equation}
где $\Phi_{d+3}$ --- ближайшее сверху число из последовательности
Фибоначчи $\{\Phi_m\}$ к числу $N+1$, и $\varphi =
\frac{1+\sqrt5}2$. Из~\cite{lh09,sp06} следует $D(N,2) = \lfloor
\log_2 N \rfloor + \lfloor \log_2(2N/3)\rfloor$. Точные или хотя
бы асимптотические оценки в замкнутой форме для $D(N,k)$ при $2 <
k < \infty$, по всей видимости, не были до сих пор получены,
несмотря на то, что, например, в работе~\cite{sp06} предложены
конструкции оптимальных схем экстремальной ширины с ограниченным
ветвлением.

Обозначим через $\alpha_k$ единственный положительный корень
многочлена $P_k(x) = {2+x+ x^2+\ldots + x^{k-2}-x^k}$. Далее будет доказана
\begin{theorem*}
$D(N,k) = \log_{\alpha_k} N \pm O(1)$.
\end{theorem*}

Легко проверить, что $\alpha_k \to \frac{1+\sqrt5}2$ при $k \to
\infty$, что согласуется с результатом~(\ref{zcg}). В частности,
из теоремы следует $D(N,3) \sim 1.65\ldots \cdot \log_2 N$,
$D(N,4) \sim 1.54\ldots \cdot \log_2 N$ и уже $D(N,9) \lesssim
1.45 \log_2 N$.

\medskip

{\bf Доказательство теоремы}

\medskip

Рассмотрим оптимальную префиксную схему глубины $D$ на $N$ входах.
Пусть ее главную цепочку образует последовательность вершин $v_0,
v_1, \ldots, v_D$, где $v_0$ совпадает со входом $x_1$,
произвольная вершина $v_d$ расположена на глубине $d$.

Вершины главной цепочки естественным образом разбивают схему на
сегменты. Если в вершинах $v_d$ и $v_{d+1}$ вычисляются суммы
$s_t$ и соответственно $s_{t+w}$, то к $d$-му сегменту отнесем
входы и вершины с индексами в интервале $[t,\, t+w]$. Параметр $w$
означает ширину сегмента. Строение сегмента оптимальной схемы
показано на рис.~\ref{pic_s} (обозначения стандартны,
см., например,~\cite{sp06,zcg06}); $h=D-d$.

\begin{figure}[htb]
\begin{center}
\begin{picture}(250,137)(0,-7)

\thinlines


\put(20,120){\vector(0,-1){120}}
\multiput(17,30)(0,25){2}{\line(1,0){6}}
\put(17,120){\line(1,0){6}}

\put(230,2.5){\vector(0,1){120}}
\multiput(227,30)(0,25){2}{\line(1,0){6}}
\put(227,2.5){\line(1,0){6}}


\multiput(81.25,2.5)(31.25,0){2}{\line(0,1){117.5}}
 \put(150,2.5){\line(0,1){117.5}}
 \put(200,30){\line(0,1){92.5}}

 \put(50,0){\line(0,1){122.5}}


\multiput(50,30)(31.25,0){3}{\circle*{3}}
 \put(200,30){\circle*{3}}
 \put(150,30){\circle*{3}}
\multiput(50,55)(150,0){2}{\circle*{3}}
\multiput(150,67.5)(50,0){2}{\circle*{3}}
\multiput(150,80)(-37.5,0){2}{\circle*{3}}
\multiput(81.25,92.5)(31.25,0){2}{\circle*{3}}

\multiput(75.25,98.5)(-5,5){4}{\circle*{3}}

\multiput(81.25,92.5)(31.25,0){2}{\line(-1,1){27.5}}
\multiput(81.25,120)(31.25,0){2}{\line(-1,0){27.5}}
 \put(150,80){\line(-5,6){33.33}}
 \put(150,120){\line(-1,0){33.33}}
 \put(200,67.5){\line(-5,6){43.75}}
 \put(200,120){\line(-1,0){43.75}}
\multiput(50,30)(31.25,0){2}{\line(1,-1){27.5}}
\multiput(50,2.5)(31.25,0){2}{\line(1,0){27.5}}
 \put(112.5,30){\line(5,-4){34.375}}
 \put(112.5,2.5){\line(1,0){34.375}}
 \put(150,30){\line(5,-3){45.83}}
 \put(150,2.5){\line(1,0){45.83}}

 \put(195.83,2.5){\line(0,-1){2.5}}


\multiput(50,54.5)(0,1){2}{\line(6,-1){150}}

\thicklines

\put(81.25,92.5){\line(-1,1){27.3}}

 \put(50,55){\line(0,-1){25}}
 \put(50,55){\line(5,-4){31.25}}
 \put(50,55){\line(5,-2){62.5}}
 \put(50,55){\line(4,-1){100}}

 \put(200,55){\line(-4,1){50}}
 \put(150,67.5){\line(-3,1){37.5}}
 \put(112.5,80){\line(-5,2){31.25}}

\multiput(81.25,30)(31.25,0){2}{\line(0,1){62.5}}
 \put(150,30){\line(0,1){50}}
 \put(200,30){\line(0,1){37.5}}


 \put(235,0.5){$\scriptstyle 0$}
 \put(235,28){$\scriptstyle h-1$}
 \put(235,53){$\scriptstyle h$}

 \put(11,118){$\scriptstyle 0$}
 \put(11,53){$\scriptstyle d$}
 \put(0,28){$\scriptstyle d\,+1$}

 \put(37,50){$v_d$} \put(197,21){$v_{d+1}$} \put(203,50){$u_{d}$}

 \put(45,126){$x_t$}  \put(190,126){$x_{t+w}$}
 \put(45,-7){$s_t$}  \put(184,-7){$s_{t+w-1}$}

 \put(70,107){$I_1$} \put(101,107){$I_2$} \put(136,103){$I_3$}  \put(183,98){$I_4$}
 \put(53,8){$Q_1$} \put(84,8){$Q_2$} \put(117,8){$Q_3$}  \put(158,8){$Q_4$}

 \put(28,75){$\ldots$} \put(208,75){$\ldots$}

\end{picture}
\caption{Строение сегмента оптимальной префиксной
схемы}\label{pic_s}
\end{center}
\end{figure}

Конструкция сегмента определяется двумя деревьями-подсхемами:
бинарным деревом, ориентированным от входов $x_{t+1}, \ldots,
x_{t+w}$ к корню $u_d$, и согласованным с ним деревом,
ориентированным от корня $v_d$ к выходам $s_t, \ldots, s_{t+w-1}$.
Ветвление вершин второго дерева ограничено числом $k$.
Согласованность деревьев означает, что второе дерево использует в
точности те интервальные суммы, которые вычисляются первым
деревом. В частности, все соседи-потомки вершины $v_d$ имеют
вторые входы строго от вершин из цепочки, соединяющей $x_{t+1}$ и
$u_d$.

Строение одного сегмента не зависит от строения других сегментов.
Поэтому максимальная ширина схемы с заданными ограничениями на
глубину и ветвление является суммой максимально возможных ширин
сегментов.

Пусть $w_k(d,h)$ означает максимальную ширину пары согласованных
деревьев, первое из которых имеет глубину $\le d$, а второе ---
глубину $\le h$, при этом вершины второго дерева имеют ветвление
не более $k$. Через $w_k(D)$ обозначим максимальную ширину
оптимальной схемы глубины $D$ с ограничением $k$ на ветвление.
Введем обозначение
\[ w^*_k(D) = \sum_{d=0}^D w_k(d,D-d). \]
Тогда получаем
\begin{equation}\label{wlow}
 w^*_k(D-1) \le w_k(D) \le w^*_k(D).
\end{equation}
Верхняя оценка описывает максимальную ширину схем, в которых для вершин $v_d$ главной цепочки
в качестве исключения допускается ветвление $k+1$. Нижняя оценка описывает ширину схем,
в которых ветвление вершин $v_d$ ограничено двойкой. В частности, $w_2(D) = w_2^*(D-1)$.


\begin{claim*}
Пусть $l=\min\{d,k-1\}$ и $d,h >0$. Тогда
\begin{equation}\label{wk}
 w_k(d,h) = \sum_{i=1}^{l-1} w_k(d-i,h-1) + 2w_k(d-l,h-1).
\end{equation}
\end{claim*}

\proof Продолжим использовать рис.~\ref{pic_s} в качестве
иллюстрации. Пусть он изображает пару согласованных деревьев $I$ и
$Q$ с корневыми вершинами $u_d$ и $v_d$ соответственно.
Непосредственные потомки вершины $v_d$ задают разбиение интервала
индексов $[t,\,t+w]$ на подынтевалы по индексам вершин цепочки,
соединяющей $x_{t+1}$ и $u_d$, и также обоих деревьев~--- на пары
согласованных поддеревьев $(I_j,\,Q_j)$. Следовательно,
\begin{equation}\label{wk0}
 w_k(d,h) = w_k(d_1,h-1) + \ldots + w_k(d_r,h-1),  \quad
 d > d_1 > d_2 > \ldots > d_{r-1} \ge d_r.
\end{equation}

При любых $d, h$ очевидно выполнено
\begin{equation}\label{w1}
 w_k(d,h) \le w_k(d+1,h). 
\end{equation}
Проверим, что при $d \ge 1$ и любом $h$ также имеет место
\begin{equation}\label{w2}
  \quad  w_k(d,h) \le 2w_k(d-1,h).
\end{equation}
Рассуждение иллюстрирует рис.~\ref{pic_iq}. Рассмотрим пару
согласованных деревьев ширины $w = w_k(d,h)$, состоящую из
бинарного дерева $I$ глубины $\le d$ и $k$-арного дерева $Q$
глубины $\le h$. Обозначим через $y$ ближайшую вершину-предок
корня $u$ дерева $I$, лежащую на пути из первого входа $x_1$.
Пусть поддерево $I_1$ с корнем в $y$ имеет ширину $\tau$. Через
$I_2$ обозначим поддерево дерева~$I$, листьями которого являются
оставшиеся $w-\tau$ входов. По свойству согласованности, во втором
дереве $Q$ имеется вершина $z$, которая является предковой в
точности для $w-\tau$ старших выходов; поддерево с корнем в
вершине $z$ обозначим через $Q_2$. Через $Q_1$ обозначим дерево,
получаемое из~$Q$ удалением поддерева~$Q_2$.

\begin{figure}[htb]
\begin{center}
\begin{picture}(247,115)(6,-5)


\thinlines


 \put(10,10){\line(1,0){60}}
 \put(10,7.5){\line(0,1){47.5}}
 \put(70,10){\line(-4,3){60}}

\multiput(42.5,68)(32.5,0){2}{\line(0,1){30}}
\multiput(42.5,68)(32.5,0){2}{\line(-1,1){27.5}}
\multiput(42.5,95.5)(32.5,0){2}{\line(-1,0){27.5}}


 \put(10,55){\circle*{3}}
 \put(75,55){\circle*{3}}
 \put(75,68){\circle*{3}}
 \put(42.5,68){\circle*{3}}
 \put(42.5,30.63){\circle*{3}}


\thicklines

 \put(75,55){\line(0,1){13}}
 \put(75,55){\line(-5,2){32.5}}

\thinlines

\multiput(42.5,12.6)(0,3){18}{\circle*{1}}


 \put(42.5,10){\line(0,-1){2.5}}
 \put(70,10){\line(0,-1){2.5}}
 \put(15,98){\line(0,-1){2.5}}

 \put(51,74){$I$} \put(25,19){$Q$} \put(46,32){$z$} \put(34,62){$y$} \put(68,47){$u$}  \put(12,57){$v$}

 \put(11,102){$x_1$}  \put(38,102){$x_{\tau}$}  \put(70,102){$x_w$}
 \put(22,100){$\cdots$} \put(52,100){$\cdots$}

 \put(6,0){$s_0$}  \put(38,0){$s_{\tau}$} \put(65,0){$s_{w-1}$}
 \put(20,-2){$\cdots$} \put(49,-2){$\cdots$}



\multiput(203,-5)(0,8){15}{\line(0,1){4}}

\thicklines

 \put(100,55){\vector(1,0){35}}

\thinlines


 \put(160,10){\line(1,0){30}}
 \put(160,7.5){\line(0,1){47.5}}
 \put(160,55){\line(4,-3){30}}
 \put(190,7.5){\line(0,1){25}}

 \put(212.5,34.4){\line(1,0){27.5}}
 \put(212.5,31.9){\line(0,1){23.1}}
 \put(240,34.4){\line(-4,3){27.5}}

\multiput(192.5,68)(52.5,0){2}{\line(0,1){30}}
\multiput(192.5,68)(52.5,0){2}{\line(-1,1){27.5}}
\multiput(192.5,95.5)(52.5,0){2}{\line(-1,0){27.5}}


 \put(160,55){\circle*{3}}
 \put(212.5,55){\circle*{3}}
 \put(245,68){\circle*{3}}
 \put(192.5,68){\circle*{3}}


 \put(240,34.4){\line(0,-1){2.5}}
\multiput(165,95.5)(52.5,0){2}{\line(0,1){2.5}}

 \put(182,83){$I_1$} \put(234,83){$I_2$} \put(169,22){$Q_1$} \put(214,38){$Q_2$}

 \put(216,56.4){$z$} \put(184,61){$y$} \put(162,57){$v$} 

 \put(159,102){$x_1$}  \put(187,102){$x_{\tau}$}  \put(205.5,102){$x_{\tau+1}$}  \put(243,102){$x_w$}
 \put(170,100){$\cdots$} \put(227.5,100){$\cdots$}

 \put(154,0){$s_0$}  \put(181,0){$s_{\tau-1}$}  \put(208,24.4){$s_{\tau}$} \put(235,24.4){$s_{w-1}$}
 \put(165,-2){$\cdots$} \put(219,22.4){$\cdots$}

\end{picture}
\caption{Преобразование пары согласованных деревьев}\label{pic_iq}
\end{center}
\end{figure}

По построению, пары деревьев $(I_1, Q_1)$ и $(I_2, Q_2)$ согласованы. Суммарная ширина пар равна $w$.
При этом глубина деревьев $I_1$, $I_2$ не превосходит $d-1$, а глубина деревьев $Q_1$, $Q_2$ не превосходит $h$.
Тем самым, (\ref{w2}) доказано.

Теперь (\ref{wk}) немедленно получается из~(\ref{wk0}) применением правил~(\ref{w1}),~(\ref{w2})
с учетом $r \le k$ и $d_r \ge 0$.
\qed

\smallskip

Переходим непосредственно к доказательству теоремы.
Оценим величину $w_k^*(D)$. Ввиду
(\ref{wk}), при $D \ge k$ имеем
\begin{equation*}
 w^*_k(D) = w_k(D,0) + \sum_{i=2}^{k-1} w^*_k(D-i) + 2w^*_k(D-k) + w_k(0,D)+\sum_{i=2}^{k-1} w_k(0,D-i).
\end{equation*}
Поскольку $w(0,h) = w(d,0) = 1$ при любых $d,h \ge 0$, получаем рекуррентное соотношение
\begin{equation*}
 w^*_k(D) = \sum_{i=2}^{k-1} w^*_k(D-i) + 2w^*_k(D-k) + k.
\end{equation*}
Указанное соотношение с учетом начальных значений $w^*_k(0),
\ldots, w^*_k(k-1)$ разрешается стандартным образом как $w^*_k(D)
\sim {c\cdot\alpha_k^D}$, где $c$~--- константа, поскольку
$\alpha_k$ имеет максимальный модуль среди корней многочлена
$P_k(x)$: действительно, модуль произвольного корня $x$
удовлетворяет неравенству
\[ |x|^k \le 2 + |x| + |x|^2 + \ldots + |x|^{k-2}, \]
откуда $|x| \le \alpha_k$. Теперь утверждение теоремы следует из
(\ref{wlow}).